# Prompt gamma ray diagnostics and enhanced hadron-therapy using neutron-free nuclear reactions

L. Giuffrida[1], D. Margarone[1], G.A.P. Cirrone[2], A. Picciotto[3] and G. Korn[1]

[1] Institute of Physics ASCR, v.v.i (FZU), ELI-Beamlines project, Prague, Czech Republic
[2] Laboratory Nazionali del Sud, INFN, Catania, Italy
[3] Micro-Nano Facility, Center for Materials and Microsystems, Fondazione Bruno Kessler, Trento, Italy

We propose a series of simulations about the potential use of Boron isotopes to trigger neutron-free (aneutronic) nuclear reactions in cancer cells through the interaction with an incoming energetic proton beam, thus resulting in the emission of characteristic prompt gamma radiation (429 keV, 718 keV and 1435 keV). Furthermore assuming that the Boron isotopes are absorbed in cancer cells, the three alpha-particles produced in each p-$^{11}$B aneutronic nuclear fusion reactions can potentially result in the enhancement of the biological dose absorbed in the tumor region since these multi-MeV alpha-particles are stopped inside the single cancer cell, thus allowing to spare the surrounding tissues. Although a similar approach based on the use of $^{11}$B nuclei has been proposed in [1], our work demonstrate, using Monte Carlo simulations, the crucial importance of the use of $^{10}$B nuclei (in a solution containing also $^{11}$B) for the generation of prompt gamma-rays, which can be applied to medical imaging. In fact, we demonstrate that the use of $^{10}$B nuclei can enhance the intensity of the 718 keV gamma-ray peak more than 30 times compared to the solution containing only $^{11}$B nuclei. A detailed explanation of the origin of the different prompt gamma-rays, as well as of their application as real-time diagnostics during a potential cancer treatment, is here discussed.

## 1. Introduction

$^{11}$B(p,α)2α nuclear-fusion was investigated for the first time in 1930s by Oliphant and Rutherford, who demonstrated that an energetic proton beam interacting with $^{11}$B nuclei, can trigger the following nuclear reaction [2]:

$$^{11}B + p \rightarrow 3\alpha + 8.7 \text{ MeV} \qquad \text{-1-}$$

Theoretical calculations, confirmed by experimental measurements have shown that the channel with highest cross-section of this reaction occurs with protons having energies around 600-700 keV. The result of such reaction main channel is the generation of three alpha-particles with typical energies between 2.5 MeV and 5.5 MeV, with a maximum at about 4.5 MeV [3, 4, 5, 6 and 7]. The main advantage of such reaction is that it does not involve neutron generation and its products (alpha particles) can be easily completely stopped in a mm-thick layer of any solid material. For these reasons the proton-Boron nuclear fusion reaction has been actively investigated by several research groups for energy production [8, 9 and 10]. In the last decade a renewed interest on this topic has been shown through the possibility to trigger such nuclear reactions by using high power pulsed laser interacting with solid B-enriched targets [11, 12, 13, 14, 15, 16, 17 and 18].

Charged energetic particles are routinely used in medicine, in particular in cancer therapy. Currently hadron-therapy is becoming one of the most important medical procedures to treat solidified tumors instead of traditional radiotherapy treatments [19, 20 and 21]. However, further improvements can be carried out not only in terms of overall cost of hadron-therapy centers, in order for them to proliferate, but also in terms of improvement of the overall treatment quality based on the tumor type and on its location in the human body. Although several candidates have been considered for this treatment technique (especially in terms of ion species), at the moment mainly protons and carbon ions have achieved very satisfactory clinical results and are routinely used for cancer treatment. The main advantage of hadron-therapy compared to traditional radiotherapy consists in the fact that charged particles release most of their energy in a few millimeters close to the end of their penetration range (so-called Bragg peak region). Such important characteristics allow to damage critically only the tumor cells and limit the interaction with the healthy tissues surrounding the tumor region. Moreover, an evident enhancement of the Relative Biological Effectiveness (RBE) has been demonstrated in the case of carbon ions compared to protons, thus resulting in a more efficient treatment for most of the radio resistant tumors [22, 23, 24 and 25]. On the other hand, by using carbon ions, issues connected with projectile and target nuclear fragmentation can arise, thus leading to unwanted dose deposition beyond the Bragg peak caused by the lightest fragments. Moreover, nuclear fragmentation of the resulting mixed field, increases the uncertainty of the biological dose, which is ultimately released into the cancer tissues.

Alternative techniques for cancer therapy are being considered in order to enhance the efficiency of such treatment. For instance, nuclear reactions involving the interaction of thermal neutrons and $^{10}$B nuclei are already used in medicine for cancer treatment in the so-called Boron Neutron Capture Therapy (BNCT) technique [26, 27 and 28]. In BNCT a solution containing $^{10}$B is injected into the human body and is absorbed in the tissues surrounding the tumor region. The used nuclear reaction is the following:

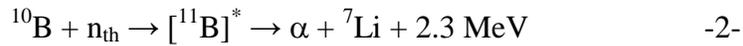

$$^{10}B + n_{th} \rightarrow [^{11}B]^* \rightarrow \alpha + {}^7Li + 2.3 \text{ MeV} \qquad -2-$$

which gives as final product an energetic alpha-particle. The produced alpha-particles are mainly localized in the tumor region due to their short propagation length and high stopping power, leading to a more efficient interaction with the tumor cells and their subsequent damage. However, unlike the classical hadron-therapy where a charged particle beam delivers a dose according to the Bragg-peak curve, in BNCT a neutron beam is slowly attenuated in the human body, thus healthy tissues lying along the neutron beam injection path are unavoidably exposed to relevant neutron doses before the final interaction of the neutrons with the boron enriched tumor cells. Furthermore, BNCT requires a so-called epithermal neutron beam which, in turn, requires sophisticated experimental devices for its generation (including shielding barriers for the patients), currently making BNCT very limited in terms of accessibility.

Recently a new approach, where the alpha-particles are generated by the interaction of a proton beam with $^{11}$B nuclei, has been proposed (not yet experimentally demonstrated) as Proton Boron Fusion Therapy (PBFT) [1]. As described in [1] in such scheme a solution containing $^{11}$B nuclei should be injected into the human body and, due to the interaction with an incoming proton, the proton-boron nuclear fusion reaction takes place and generates three alpha-particles, which can potentially destroy cancer cells more efficiently compared to a conventional proton therapy treatment. Moreover, the authors of [1] propose the generation and measurement of characteristic gamma-rays at 718 keV as real-time imaging technique. In the following we will firstly demonstrate that the proton-Boron nuclear fusion reaction using $^{11}$B nuclei does not produce prompt gamma rays and, as a consequence, does not have an "online" imaging capability. However, we will show that the combination of $^{11}$B and $^{10}$B can be used for this purpose. Moreover, depending on the tumor type and location, the concentration of Boron atoms in the solution, as well as the relative concentration of $^{11}$B and $^{10}$B isotopes, has to be optimized. In fact, while tumors with larger size, more resistant and closer to sensitive tissues in the human body would benefit from a higher concentration of $^{11}$B compared to $^{10}$B nuclei, smaller, non-radiation resistant, deeper-seated tumors would need a higher concentration of $^{10}$B compared to $^{11}$B in order to compensate the higher absorption in the human body of the characteristic prompt gamma radiation, ultimately to be used for real-time imaging.

Furthermore, the source of the physical dose enhancement reported in [1], identified as main feature of PBFT, is incorrect. In fact, as it will be discussed in detail below, the treatment enhancement reported in [1] cannot be ascribed to a relevant increase of the physical dose (depicted in Fig.2 and Fig.3 of [1]) and, in addition, there is no prompt gamma radiation generated in the p-$^{11}$B fusion nuclear reaction at 718 keV claimed in [1]. Furthermore, while in [1] only the use of the proton-boron nuclear fusion reaction is discussed, in the present work additional nuclear reactions, occurring due to the presence of $^{10}$B nuclei, are proposed for the generation of characteristic prompt gamma-rays.

An innovative scheme for simultaneous prompt gamma ray imaging and enhanced hadron-therapy using neutron-free nuclear reactions is presented and discussed in this work through the use of Monte Carlo simulation outputs.

## 2. Methods

The interaction of protons with $^{10}$B nuclei triggers aneutronic nuclear reactions generating characteristic prompt gamma-rays which can be used for a real-time monitoring of the treatment and potentially for a sort of dynamic treatment with a feedback control based on real-time dose measurement.

The characteristic prompt gamma-ray peaks due to the interaction of the $^{10}$B nuclei with the incoming energetic protons are peaked at 429 keV, 718 keV and at 1435 keV. The peak at 429 keV is ascribed to the $^{10}$B (p, α) $^{7}$Be nuclear reaction. The peak at 718 keV is mainly ascribed to the inelastic scattering of the proton, $^{10}$B ($^{10}$B* (p,p`) $^{10}$B), but it can also be produced from the $^{10}$B (p,n) $^{10}$C reaction and the consequent $^{10}$C β+ decay but with a much less cross section (see Fig.3) into the $^{10}$B* (not resulting in a prompt gamma ray emission). The peak at 1435 keV is due to the $^{10}$B (p,p`) $^{10}$B* nuclear reaction which is generated when the $^{10}$B* nuclei decay from the 2.15 MeV level to the 718 keV level. All these peaks (excluding the 718 keV produced in the $^{10}$C β+ decay) are prompt gamma-rays which can be used for real-time measurements.

The $^{11}$B nuclei interacting with protons can trigger the $^{11}$B (p,2n) $^{10}$C nuclear reaction β+ decaying in $^{10}$B* and emitting a gamma-ray at 718 keV, but the cross-section of such reaction is very low compared to the cross-sections of the other reactions with $^{10}$B nuclei above described. Moreover the gamma-rays produced in such reaction are not prompt, thus the specific reaction is not useful for a real-time measurement. Additional details about these reactions will be given in the section "Results and discussions" and in Fig.3.

The proposed treatment procedure benefits from both proton therapy, since protons are mainly used as projectiles to trigger the nuclear fusion reaction, and heavier ion therapy, since alpha-particles generated in the nuclear fusion reactions have a higher LET (Linear Energy Transfer) and cause more efficient damages in the single cell.

It is worth highlighting that the presence of an optimized mixture of $^{10}$B and $^{11}$B is crucial for the method hereby described. In fact, the $^{11}$B nuclei are important for triggering the p-$^{11}$B nuclear fusion reaction (enhanced treatment capability), while the $^{10}$B nuclei are important for the generation of prompt gamma-rays (real-time imaging capability), thus allowing to carry out a potential dynamic treatment. Moreover, in order to maximize the effects of the cancer treatment and the imaging capability, the optimization of the ratio $^{10}$B/$^{11}$B is crucial. On the one hand, for superficial tumors it is more convenient to increase the $^{10}$B concentration with respect to the $^{11}$B one in order to maximize the production of gamma prompt peaks. The presence of such exclusive peaks, will improve the quality of the beam imaging process. On the other hand, for tumor depth larger than 10 cm of soft tissue, the gamma attenuation will reduce the total amount of detectable gammas, thus it will be more convenient to increase the percentage of $^{11}$B with respect to $^{10}$B.

Fig.1a shows a conceptual sketch of the system used for a potential simultaneous cancer treatment and diagnostic approach (gamma-ray detector, proton beam, patient positioning system). The same figure shows the geometry used for our numerical simulation performed by using the MCNPX 2.7 Monte Carlo code [29], along with details of the two different planes, xy in b) and xz in c).

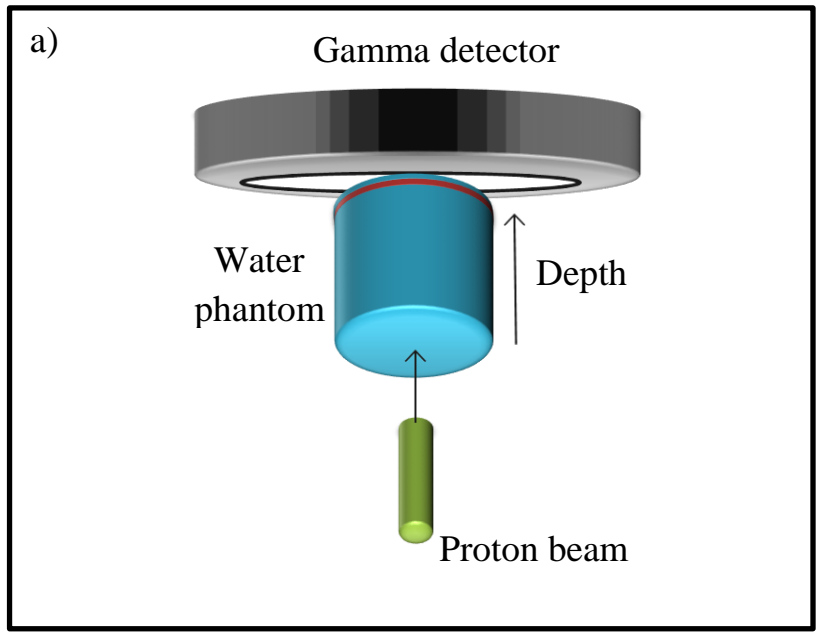

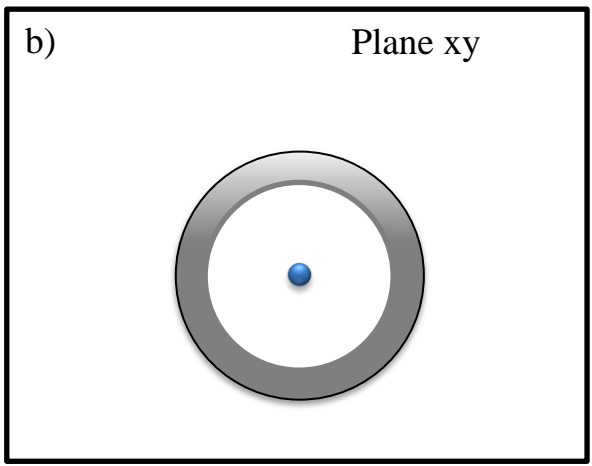 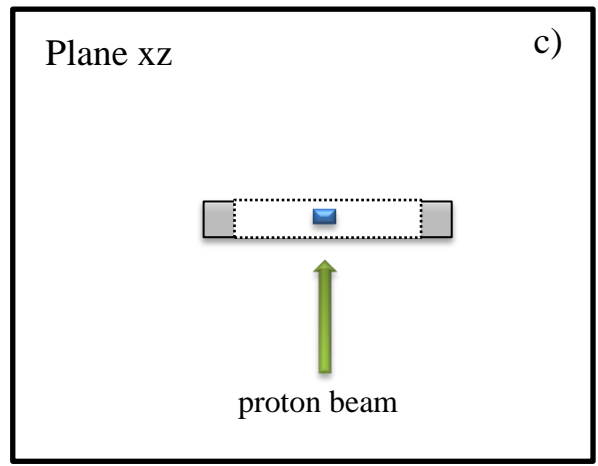

**Fig.1**: setup employed for MCNPX simulations. a) is a sketch of the overall system. b) and c) are respectively, sketches showing the xy and xz planes. The cylindrical shell in gray represents the Ge gamma-rays detector, the light blue shape is the water phantom including the B isotopes doped region (in red). The incoming proton beam is represented in green.

A water cylinder (here labelled as H2O) with density of 1.01 g/cm$^3$, with a total length of 3.5 cm and with a diameter of 6 cm (simulating the human body) is irradiated by a mono-energetic proton beam with energy of 60 MeV, placed along the z axis. In order to simulate the cancer region surrounded by the solution containing B atoms, a small cylinder (3 mm long and 4 cm wide) is placed in the Bragg peak region in the position between 2.9 and 3.2 cm within the water cylinder. As real-time diagnostics a Ge gamma-ray cylindrical-shell detector with internal diameter of 26 cm and thickness of 4 cm is used. The Ge detector is placed around the water phantom covering the whole sample in order to maximize the signal coming out of the irradiated sample.

Gamma ray characteristic emission peaks have been studied in different conditions of Boron concentration. In particular, we studied three 'pure' configurations where only $^{10}$B, $^{11}$B or natural B have been considered (here respectively labelled as B10, B11 and B) and one water-based solution containing 1% of $^{11}$B, here labelled as B11(1%). The isotopic composition of the natural Boron is $^{10}$B at 20% and $^{11}$B at 80%. We have also studied a configuration containing only water (in the text labelled as H2O).

Our numerical investigation allowed explaining the origin of all characteristic gamma-ray peaks in the energy range 200 keV - 1.5 MeV. For our simulations the "ftally 8" (energy distribution of pulses created in the detector by radiation) has been used. In the second part of the simulation study, the dose released by the energetic proton beam in H2O was also calculated to investigate how the Bragg peak can change intensity and position within the sample, according to different concentrations of $^{11}$B within this region. In order to perform such simulation the "ftally 6" (energy deposition averaged over a cell given in MeV/g) has been used. In this case H2O was divided in small cylinders with length of 0.1 mm each. The small $^{11}$B cylinders had the same length of the water cylinders (0.1 mm) and 20 of them were positioned in the Bragg peak region. The proton and alpha particle energy deposition in H2O was also studied by using the "tmesh 3" (MeV/cm$^3$/source particles). The "tmesh 1" (flux given in number of particles per cm$^2$) was used to calculate the alpha-particle and proton distributions through the depth of the sample. A very high statistics of injected protons ($6*10^8$ proton/simulation) has been used for each of these simulations. Default dataset for the cross-sections given by MCNPX were kept without changes to avoid unrealistic results arising from potential errors in the simulation modelling. For more detailed information about the meaning of ftallyes and tmeshes one can refer to the MCNPX 2.7.0 manual.

## 3. Results and discussions

The first investigation performed through MCNPX simulations provides a preliminary idea of the characteristic gamma-rays which could come out of the human body during a potential cancer treatment. In these preliminary simulations gamma-ray spectra up to 1.5 MeV with a resolution in energy of less than 1 keV are obtained. A particular attention is given to the following gamma-ray peaks: 429 keV, 718 keV and 1435 keV. A Ge gamma-ray cylindrical-shell detector is used in the simulations in order to maximize the signal generated from the irradiated samples. The different samples ($H_2O$, B10, B11 and B) are irradiated by a mono-energetic proton beam of 60 MeV. Results of the emitted gamma-rays with different samples are shown in Fig.2 a) for $H_2O$ and B10 (on the top of the figure), B11 (on the middle of the figure) and B (on the bottom of the figure), respectively.

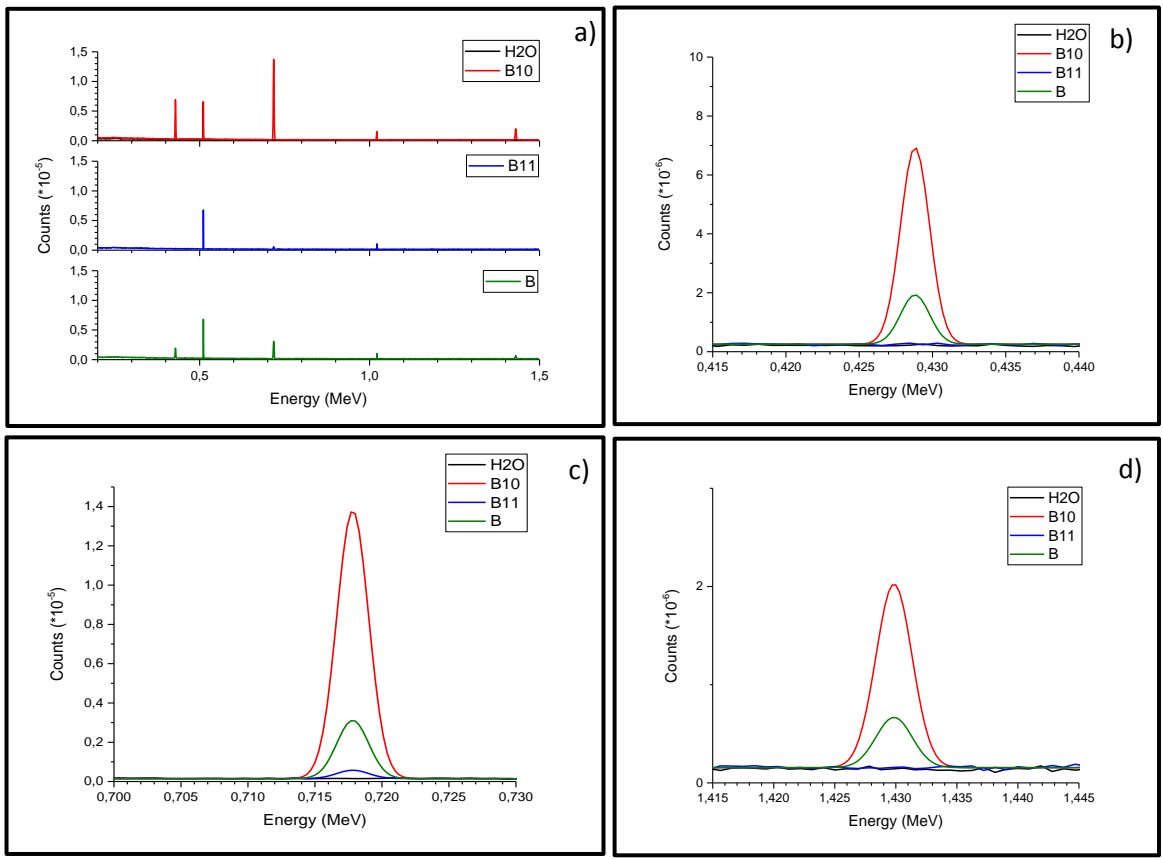

**Fig.2**: a) gamma-ray spectra emitted in the case of the sample $H_2O$ and B10 (on the top of a), 11B (on the middle of a) and B (on the bottom of a). Detailed comparison of the gamma-ray peaks at 429 keV b), at 718 keV c) and at 1435 keV d) are also here showed.

The first important result is evident from the H2O spectrum where there is no presence of characteristic gamma-ray peaks concerning the mentioned above. The 511 keV peak comes from the positron annihilation followed to the electron-positron pair production in the target material and in the detector itself. In the case of B10 (red curve) there are three different clear prompt gamma-ray peaks: 718 keV (the most intense), 429 keV and 1435 keV. A completely different spectrum is reported in the middle part of Fig.2a for the gamma-ray emission from B11 (blue curve). Here only the peak at 718 keV is identified, however the relative peak intensity is much lower compared to B10 (more than a factor 30). Substantial differences in the gamma-ray spectrum features are shown when comparing the case of B10 and the case of B11. In the latter case there is no presence of gamma peaks at 429 keV and at 1435 keV, only the peak at 718 keV appears, although the relative intensity is very low compared to B10. This is an important point since the presence of both boron isotopes ($^{11}$B and $^{10}$B) in an optimized relative percentage (depending on the tumor) is necessary for the proposed simultaneous imaging and treatment technique, differently than what stated in [1] where only $^{11}$B nuclei are used for the proposed gamma-ray imaging.

Fig.2a (bottom part) shows the results obtained with B (green curve). These results are in agreement with what discussed above in terms of gamma ray peaks at 429 keV, 718 keV and 1435 keV, which are present with different relative intensities, thus showing an intermediate behavior compared to B11 and B10. Fig.2 shows also the comparisons among the gamma-ray peaks at 429 keV (b), 718 keV (c) and 1435 keV (d) for B10 (in red), B11 (in black) and B (in green), respectively. The same comparison is reported in Table I, where the peak values are normalized to the 718 keV peak intensity recorded in the case of the B10 sample.

Several important conclusions can be anticipated from a first comparison among the different simulated spectra. All the above described characteristic gamma-ray peaks are mainly ascribable to the presence of $^{10}$B nuclei. The major peak at 718 keV is present for all the different simulated samples, however in the case of B11 the intensity of the gamma-ray signal is more than 30 times lower than in the case of B10. From Table I it is also clear that in the case of B (containing 20% of $^{10}$B nuclei) the gamma-ray peak at 718 keV is about 7 times more intense than the B11 case. This is a very important result because it shows that it is possible to modify and optimize the intensity of the detected gamma-ray peaks by changing the relative concentration of $^{10}$B and $^{11}$B nuclei in the injected solution. The minor gamma-ray peaks (429 keV and 1435 keV) are present only in the cases of B10 and B.

| Target | Intensity at 429 keV | Intensity at 718 keV | Intensity at 1435 keV |
|---|---|---|---|
| B | 12.4 | 22 | 3.8 |
| $^{10}$B | 49.6 | 100 | 13.8 |
| $^{11}$B | NA | 3.1 | NA |

**Table I**: normalized gamma-ray peaks intensity at 429 keV, 718 keV and 1435 keV for the different investigated cases (B, B10 and B11). Here values are normalized respect to the value of the 718 keV gamma-rays peak in the sample B10.

Beside the outputs of our numerical simulations in terms of gamma-ray spectral line intensities, it is important to explain the origin of the above mentioned characteristic gamma ray peaks. In the case of B10 the following reaction is responsible of the emission of gamma-rays at 718 keV:

$$^{10}B\ (p,p`\gamma)\ ^{10}B \qquad\qquad -3-$$

This is an aneutronic nuclear reaction, with a non-negligible cross-section for energies of a few MeV, based on the inelastic scattering of the $^{10}$B nuclei with the incoming protons [30].
The following nuclear reactions are also possible: $^{10}$B (p,n) $^{10}$C and $^{10}$C β+ decay, evolving into a $^{10}$B* exited state and then emitting a gamma-ray at 718 keV [30 and 31]. However, only the inelastic reaction -3- leads to a prompt gamma-ray peak, while the other reactions do not generate prompt gamma radiation.
Furthermore, the interaction of the protons with $^{10}$B nuclei can also trigger another nuclear reaction:

$$^{10}B\ (p,\alpha)\ ^{7}Be \qquad\qquad -4-$$

where the final result is a prompt gamma-ray emission at 429 keV [32].

For a full explanation of the gamma spectrum obtained in the simulations for the B11 sample, the following series of nuclear reactions have been identified: $^{11}$B (p,2n) $^{10}$C, followed by a β+ decay of $^{10}$C, thus populating the $^{10}$B* exited state and finally emitting a gamma-ray at 718 keV [33]. It is worth mentioning that the cross-section of $^{11}$B (p,2n) $^{10}$C is negligible compared to $^{10}$B (p,n)$^{10}$C cross section for a proton energy of a few MeV. This can explain the large difference in terms of peak intensity for the 718 keV gamma radiation shown as a result of our numerical simulations (30 times lower for B11 compared to B10). In conclusion, the above identified nuclear reactions can fully explain the origin of the gamma-ray peaks shown by our Monte Carlo simulations.

Additional details are shown in Fig.3 where the rates of the 718 keV gamma-ray peaks for different reactions are compared. The relative reaction rate for the inelastic scattering and for the $^{10}$B (p,n) $^{10}$C nuclear reaction are compared. In the area of interest within the Bragg peak (represented with a blue dotted region in Fig. 3) the relative rate of $^{10}$B (p,n) $^{10}$C is less than 10% compared to the inelastic scattering. This clearly explains that the main contribution of the 718 keV peak is ascribable to the $^{10}$B (p,p`γ) $^{10}$B inelastic scattering.

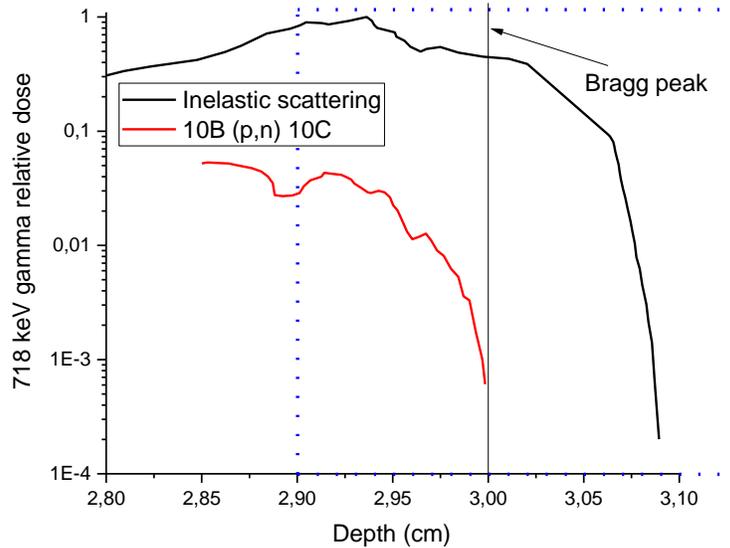

**Fig.3**: comparison of the 718 keV characteristic gamma-ray generation rate for different nuclear reactions: $^{10}$B (p,p`) $^{10}$B* (in black) and $^{10}$B (p,n) $^{10}$C (in red).

The second part of this numerical work supports the proposed new tumor treatment method. This study provides a different understanding of such approach with respect to the interpretation described in [1]. The H2O sample, irradiated with an incoming mono-energetic proton beam of 60 MeV, is used as a reference. A preliminary study allows estimating the dose released by the protons in H2O, which shows the Bragg peak location at 3 cm from the sample surface. The reference sample is compared with the B11(1%) one, which is a realistic case considering the typical concentration of Boron used in the BNCT therapy. The comparison between H2O (in black) and B11(1%) (in red) simulation outputs is shown in Fig.4a.

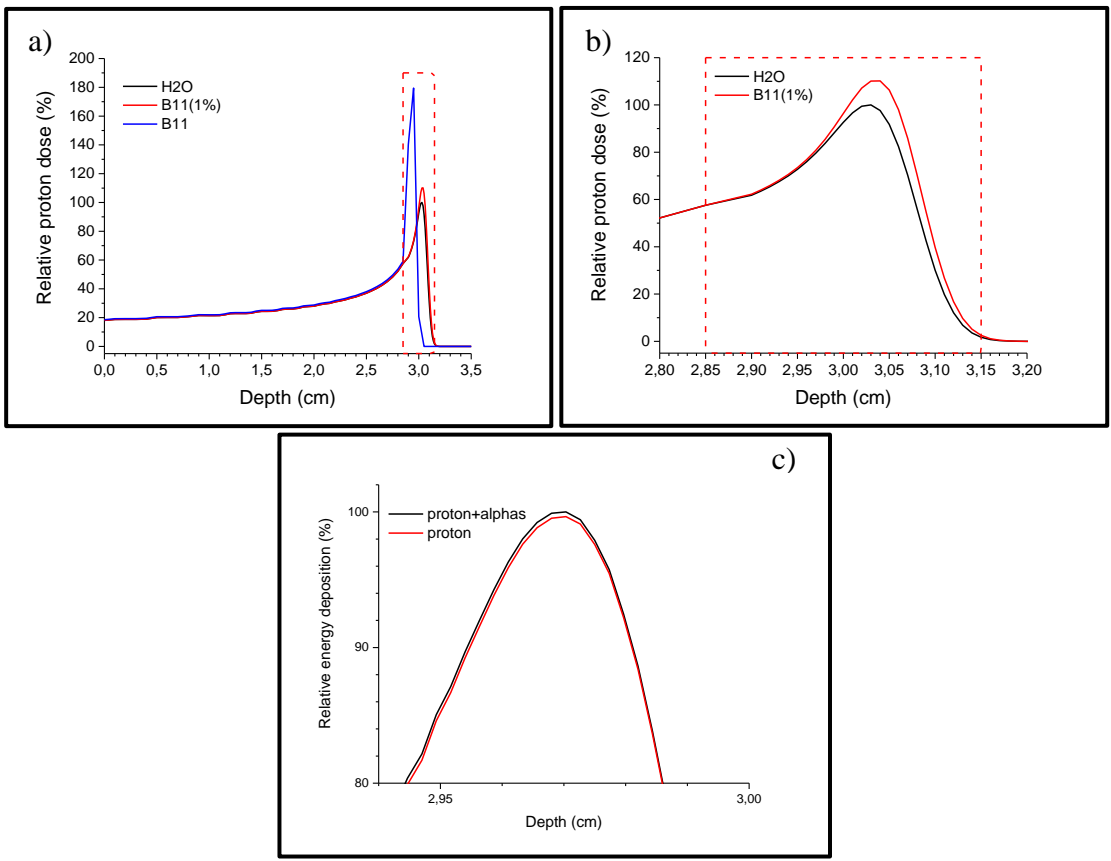

**Fig.4**: relative dose released by protons within H2O (black line), B11(1%) (red line) and B11 (blue line) (a). The red dashed region represents the area where the [11]B nuclei are located within the doped water phantoms. b) is a zoom in the depth range between 2.8 cm and 3.2 cm, showing a detail of the relative dose released by protons for H2O and B11(1%). Fig. 4 c) shows the relative energy deposited in B11(1%) by protons and alphas (black line) and only by protons (red line).

Here only the proton dose without the potential alpha-particle contribution is considered. The calculated values are normalized considering the H2O results as a reference case (100% intensity at the peak). A small enhancement (less than 10%) of the physical dose released within the Bragg peak region in the case of B11(1%) is marked in red. The position of the Bragg peak in both cases is practically the same. This result shows that the small enhancement in the physical dose observed in the case of B11(1%) is not ascribable to the alpha particles produced in the proton-Boron nuclear reaction, contrary to what reported in [1], however it is mainly due to the presence of the $^{11}$B atoms, which changes the composition and the density of the water phantom. In order to further demonstrate the previous statement, we have considered an unrealistic case where the concentration of $^{11}$B nuclei is 100% (overestimation) and the Boron density is 2.46 g/cm$^3$. The simulation results are shown in Fig.4a (the blue line shows the Bragg curve). A large enhancement of the Bragg peak compared to the H2O case is shown (from 100% to 180%), clearly demonstrating that this change is ascribable to the different density of the irradiated sample.

Fig.4c shows the relative energy deposition for protons (in red) and for protons + alpha-particles (in black) for the B11(1%) sample. The comparison shows that the deposited energy is coming from the protons since the two curves are practically the same (only 0.3% of the total contribution is ascribable to the alpha-particles). In order to support this idea, a second set of simulations has been carried out (see Fig.5), where the flux (number/cm$^2$) of the generated alpha-particles (a), and protons (b), within the sample B11(1%) is estimated. In Fig. 5a the alpha-particles yield (red line) is shown in comparison to the dose absorbed in the whole sample (black line). A Gaussian fit of the alpha-particles distribution was carried out (dashed black line). The low FHWM of the curve (1.3 mm) leads to the conclusion that the alpha-particles are confined in a very well localized region, thus allowing to define very precisely the position where they release the dose during the treatment. Another important point is that the peak of the alpha-particles is placed approximately in the same position where the protons release their maximum dose (i.e. around the Bragg peak). Thus, the exact knowledge of the Bragg peak position and of the position of the peak of maximum alpha-particle dose release can be used to enhance the efficacy of the treatment. Fig.5b shows the number of protons passing through the irradiated sample. This plot is in agreement with the one reported in Fig.4a, showing that the amount of protons starts to decrease in the Bragg peak region, falling down to zero after a few mm. It is important to note that the relative amount of alpha-particles (Fig. 5a) compared to protons (Fig. 5b) is much lower (around 5 orders of magnitude). This again means that the physical dose enhancement close to the Bragg peak (around 10%) is mainly due to the protons and only in minor amount to the alpha-particles generated in the nuclear fusion reaction, differently than stated in [1]. We anticipate a potential higher efficacy of the treatment due to the enhancement of the biological damage (not mentioned in [1]) based on the fact that the alpha-particles generated by the p-$^{11}$B nuclear fusion reaction practically lose all their energy (and are stopped) within the single cell and, moreover, have a higher LET leading to the generation of more complex damages inside the cell compared to protons used in conventional proton-therapy.

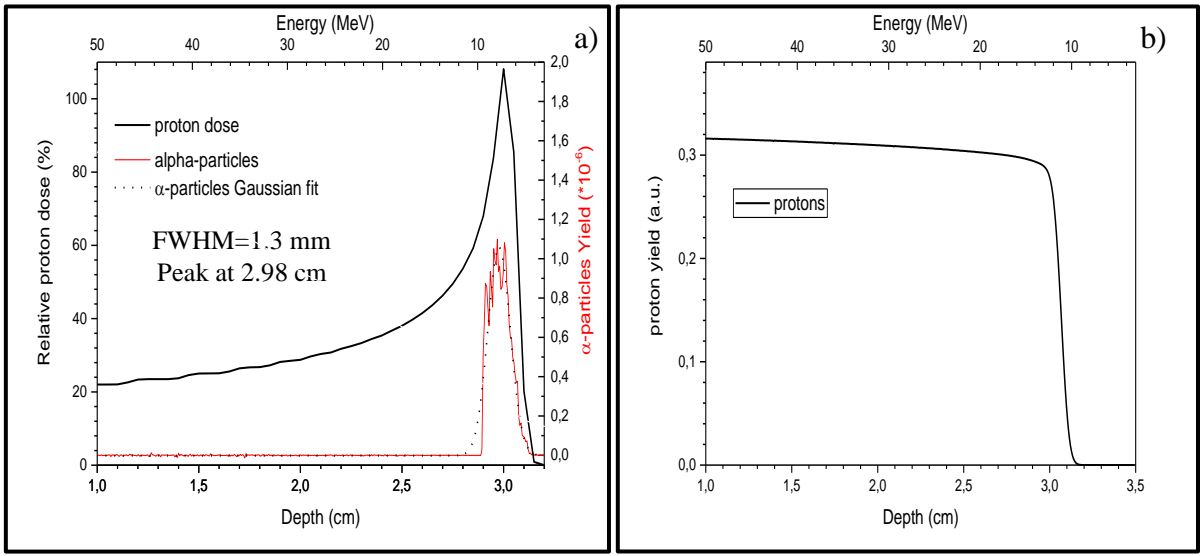

**Fig.5**: alpha-particle a) and proton b) yield within the 11B(1%) sample. In a) the black curve is the proton dose released in the whole sample while the black dashed line is a Gaussian fit of the alpha-particle distribution (in red) showing the FWHM of 1.3 mm and the peak at 2.98 cm.

Moreover, although the standard proton-therapy is very efficient, a small part of the proton dose is in any case released to the healthy tissues surrounding the cancer cells. The approach proposed in this work is potentially more precise. In fact, a rough estimation indicates that $^{11}$B nuclei injected into the human body and located in the cancer cells, generate alpha-particles with energy of 2-5 MeV, thus having a range of penetration in the cells of less than 20 micrometers. Since the typical size of cancer cells is around 20-30 microns, most of the alpha-particles release their dose inside the cells, thus enhancing the possibility to destroy only the tumor tissues and to spare the healthy ones. In conclusion, our numerical investigations show that the alpha-particles coming from the p-$^{11}$B aneutronic nuclear fusion reaction are generated in a very localized region corresponding to the maximum alpha-particle flux, thus allowing the possibility to irradiate the cancer region additionally in a very precise way, basically without losses outside the cancer region, thus avoiding damages in healthy tissues surrounding the tumor.

## 4. Conclusions

The possibility to inject into the human body a solution containing an optimized combination of $^{11}$B and $^{10}$B, which interacting with energetic protons can trigger proton boron neutron-free nuclear fusion reactions, has been investigated in this paper. This technique allows performing simultaneous real-time gamma-ray imaging and enhanced cancer treatment, thus having a very strong impact in medicine and in particular in cancer therapy, hence adding new features to the well-known and well-established proton-therapy.

Monte Carlo simulations allowed understanding the origin of the characteristic gamma-ray line emission from the irradiated samples, which can be used for a real-time imaging of the treatment. We have pointed out that the presence of $^{10}$B nuclei is mandatory for the emission of such characteristic prompt gamma-rays (i.e. 429 keV, 718 keV and 1435 keV) and that the gamma-ray peak at 718 keV, ascribed to the p-$^{11}$B fusion reaction in [1], is produced by a different nuclear reaction ($^{11}$B (p,2n) $^{10}$C, followed by the β+ decay of 10C, then evolving into the $^{10}$B* exited state). This reaction does not produce prompt gamma-rays and, as a consequence, cannot be used as a potential online imaging technique, differently than claimed in [1]. Furthermore we demonstrated that the intensity of the 718 keV gamma-ray peak is enhanced more than 30 times when using $^{10}$B nuclei instead of $^{11}$B nuclei.

In conclusion, our numerical simulations demonstrate that the enhancement of the dose reported in [1] is not due to the occurrence of the p-$^{11}$B fusion reaction but it is ascribable to the change in the density of the sample containing dopant nuclei. However, although the enhancement of the physical dose in the doped sample is negligible, if compared with a reference sample, we expect an increase of the biological dose due to the fact that the alpha-particles generated by the p-$^{11}$B nuclear reaction in the cancer cells lose most of their energy inside the cells themselves and have a higher LET compared to protons.

For a deeper understanding of the proposed approach we are planning to perform an experimental campaign using a conventional accelerator which would involve both the investigation of prompt gamma-ray emission and the study of biological dose enhancement in cancer cells.

**Acknowledgments**

This work has been supported by the project ELI - Extreme Light Infrastructure – phase 2 (CZ.02.1.01/0.0/0.0/15_008/0000162 ) from European Regional Development Fund.
This research was also sponsored by the Czech Science Foundation (project No. 15-02964S).